\documentclass[11pt]{article}
\usepackage{blois2002,epsfig}
\def\be{\begin{equation}}
\def\ee{\end{equation}}
\def\bea{\begin{eqnarray}}
\def\eea{\end{eqnarray}}

\begin{document}
\vspace*{4cm}
\title{ELECTROWEAK PENGUINS AND LARGE $N_c$}

\author{S. PERIS }

\address{Grup de Fisica Teorica and IFAE, Univ. Autonoma de Barcelona\\
08193 Barcelona, Spain}

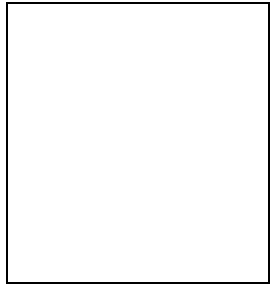
\maketitle\abstracts{ I review the progress made in the
calculation of the electroweak penguins $Q_{7,8}$ in the
framework of the Hadronic Approximation to large-$N_c$ QCD, and
give an update of results.}

A nonzero value of $\epsilon'/\epsilon$ signals the existence of
direct CP violation in the decay $K\rightarrow \pi\pi$ and is a
measure of the amount of CP violation in the Standard Model. At
present the world average is given by~\cite{Nakada}
\begin{equation}\label{nakada}
  \frac{\epsilon'}{\epsilon}= (16.6 \pm 1.6) \times 10^{-4}\  .
\end{equation}
To compute this number in the SM has become a major theoretical
challenge~\cite{Martinelli}. Roughly~\cite{Buras},
$\epsilon'/\epsilon$ is dominated by the difference of
$K\rightarrow \pi\pi$ matrix elements of the $Q_6$  and $Q_8$
operators,
\begin{equation}\label{Q6}
  Q_{\{6,8\}}= - \sum_{q=u,d,s}\{8,12\ e_q\}\  \left(\overline{s}_L
q_R\right)\left(\overline{q}_R
  d_L\right)\ ,
\end{equation}
where the factors $e_q=(2e/3,-e/3,-e/3)$ are the quark electric
charges. These factors make a profound difference between $Q_6$
and $Q_8$ at the meson level as they change their chiral
properties making $Q_8$ an operator which starts at
$\mathcal{O}(p^0)$, whereas $Q_6$ only starts at
$\mathcal{O}(p^2)$. This makes $Q_8$ a simpler object to deal
with and allows a connection to the well-known problem of the
$\pi^{+}\!-\pi^{0}$ electromagnetic mass difference.

Strong interactions make these operators change with the
renormalization scale and mix with others so that, strictly
speaking, they can never be considered in isolation. In
particular $Q_8$ mixes with $Q_7$,
\begin{equation}\label{Q7}
   Q_{7}= 6 \sum_{q=u,d,s} \ e_q \left(\overline{s}_L \gamma^{\mu}
d_L\right)\left(\overline{q}_R \gamma_{\mu} q_R\right)\ .
\end{equation}
This apparent complication will actually be very helpful in the
calculation of $Q_8$.

The main effect is due to the fact that $Q_7$ and $Q_8$ contribute
to the operator
\begin{equation}\label{operator}
  \mathrm{Tr}\left(U\lambda_L^{(23)}U^\dag Q_R\right)\ ,
\end{equation}
where $(\lambda^{(32)}_L)_{ij}=\delta_{i3}\delta_{j2}$ and
$Q_R=\mathrm{diag}(2/3,-1/3,-1/3)$. This operator is a
``flavor-rotated'' version of the operator responsible for the
$\pi^+-\pi^0$ electromagnetic mass difference, which is a problem
that has been reasonably understood for a long time~\cite{pion}.
This relationship is the underlying reason why both the physics
of the $\pi^+-\pi^0$ mass difference and the bosonization of
$Q_{7,8}$ into the operator (\ref{operator}) are governed by one
and the same Green's function, which turns out to be the analog
of the vacuum polarization but between the left- and right-handed
currents $\overline{q}_{L,R}\gamma^{\mu} q_{L,R}$. Furthermore,
$Q_7$ can be looked upon as due to the exchange of an imaginary
``photon'' with couplings to the quarks which are nondiagonal in
flavor and whose ``propagator'' is unity instead of the usual
$1/q^2$. One then obtains that~\cite{Knecht}
\begin{equation}\label{Q7bos}
  Q_7=6 <O_1(\mu)>\ \mathrm{Tr}\left(U\lambda_L^{(23)}U^\dag
  Q_R\right)^\dag\ ,
\end{equation}
with
\begin{equation}\label{O1}
  <O_1(\mu)>=<\overline{s}_L \gamma^{\mu}d_L\ \overline{d}_R
  \gamma_{\mu}s_R>=\frac{1}{6}\left(-3ig_{\mu\nu} \int\frac {d^4q}{(2\pi)^4}
\Pi_{LR}^{\mu\nu}(q)\right)_{\overline{MS}}\ ,
\end{equation}
where current conservation implies $\Pi_{LR}^{\mu\nu}(q)= (q^{\mu}
q^{\nu}- q^2 g^{\mu\nu}) \Pi_{LR}(q^2)$, since we are in the
chiral limit. Because our imaginary photon has unity as its
propagator, this integral is divergent and has to be regularized
and renormalized. The subscript $\overline{MS}$ refers to this
fact. Eq. (\ref{O1}) shows an example of how a coupling constant
in the effective meson Lagrangian (i.e. $< O_1(\mu)>$) is related
to integrals over euclidean momentum of QCD Green's functions.
This is also true in general.

However, in order to really compute $<O_1>$ one still needs to
know the function $\Pi_{LR}(q^2)$ for all values of $q^2$ which is
not known. The large-$N_c$ limit of QCD simplifies the problem
because it determines the analytic structure of this function: it
has to be meromorphic. Therefore it can only have poles (and no
cut), which in physical terms correspond to the meson states.
However this is still not enough as the number of poles is
infinite and, since the solution to large-$N_c$ QCD has not been
found, the location of these poles and their residues are
unknown. It is at this point that a further approximation has to
be made. This approximation, which we name ``The Hadronic
Approximation''~\cite{HA} to large-$N_c$ QCD, is defined by
keeping only a \emph{finite} number of resonances, whose residues
and masses are fixed by matching to the first few terms of both
the chiral and the OPE expansions~\footnote{In the past we used
the term ``Minimal Hadronic Approximation'' for cases in which
only the leading term in the OPE expansion was used in the
matching.} of $\Pi_{LR}(Q^2)$\ \footnote{Notice that this is
possible because $\Pi_{LR}(Q^2)$ has no parton-model $\log Q^2$;
it would not be possible in the case of the vector-vector
correlator, for instance.}. This is called in mathematics a
rational approximant and is an interpolating function, ratio of
two polynomials, which by construction has the same low- and
high-$Q^2$ behavior as the full $\Pi_{LR}(Q^2)$ of large-$N_c$
QCD. This approximation rationalizes and systematizes old
phenomenological approaches such as Vector Meson Dominance;
incorporating chiral symmetry (to control long distances) and the
OPE (to control short distances). It is also an improvement over
approaches where only the OPE constraints were considered, such
as QCD sum rules, because it also uses chiral symmetry at long
distances.

An analysis of Aleph data~\cite{Boris} shows that just the pion
and one vector and one axial-vector states do a pretty good job
provided their masses and decay constants fulfill the
above-mentioned long and short-distance constraints. I urge the
reader to take a look at the curves in Fig. 1 of the first paper
in  Ref. ~\cite{HA}. In this case one obtains the remarkably
simple expression
\begin{equation}\label{PiMHA}
  -Q^2 \Pi_{LR}(Q^2) =\frac{f_{0}^2 M_V^2 M_A^2}{(Q^2+M_V^2)
  (Q^2+M_A^2)}\ ,
\end{equation}
where $f_{0}=87\pm 3\ {\rm MeV}$, $M_V=748\pm 29\ {\rm MeV}$ and
$M_V^2/M_A^2 \simeq 0.50\pm 0.06$~\cite{Boris}. These values do
not have to be exactly equal to the physical ones although it is
natural to expect that, if the Hadronic Approximation and
large-$N_c$ work, they have to be close to each other.

Having the function $\Pi_{LR}(Q^2)$, one can plug it in Eq.
(\ref{O1}) and compute the integral. The result is
\begin{equation}\label{O1result}
  <O_1>(\mu)=-\ \frac{3}{32 \pi^2}\ \ \frac{f_0^2 M_V^4}{1-
  \frac{M_V^2}{M_A^2}}\ \ \log\left[\frac{M_V^2}{M_A^2}
  \left(\frac{\Lambda^2}{M_A^2}
  \right)^{\frac{M_A^2}{M_V^2}-1}\right]\ ,
\end{equation}
with $\Lambda^2\equiv \mu^2 \exp(1/3 + \kappa)$, where
$\kappa=-1/2(3/2)$ in NDR(HV) schemes,  respectively. The control
over the renormalization scheme ($\gamma_5$, evanescent
operators, etc...) is the result of the OPE constraints used in
the construction of the Hadronic Approximation~\cite{Knecht}. It
is obvious that knowing $<O_1>$ is tantamount to knowing any
matrix element like, e.g., $<\pi\pi\mid Q_7  \mid K>$: one only
expands the $U$ in Eq. (\ref{Q7bos}).

The calculation of the matrix elements of $Q_8$ is a bit more
tricky because it is the product of two scalar-pseudoscalar
densities -see Eq. (\ref{Q6})-. The previous analogy with an
imaginary ``photon''\footnote{This ``photon'' is now even more
strange: it doesn't have spin one!.} still works and leads to a
result similar to (\ref{Q7bos}),(\ref{O1}) but involving these
scalar-pseudoscalar densities, to wit~\cite{Knecht}
\begin{equation}\label{Q8bos}
  Q_8=- 12  <O_2(\mu)>\ \mathrm{Tr}\left(U\lambda_L^{(23)}U^\dag
  Q_R\right)^\dag\ ,
\end{equation}
where
\begin{eqnarray}\label{O2}
  <O_2(\mu)>&=&\ <\overline{s}_L s_R\ \overline{d}_R d_L>
  \nonumber \\
  &=&\frac{1}{4} <\overline{\psi} \psi>_{\overline{MS}}^2 + \left(\int \! \! \!
  \frac{d^D q}{(2\pi)^D}\int\!\! d^4 x\  e^{i qx}\ \langle 0\vert
T\left[\bar{d}_L d_R(x)\,\, \bar{s}_R s_L(0)\right]|0
\rangle_{C}\right)_{\overline{MS}} \ .
\end{eqnarray}
Taking the large-$N_c$ limit selects the quark condensate in the
previous equation. However, there is some evidence -although
circumstantial~\cite{Moussallam}- that this limit in the
scalar-pseudoscalar sector may have somewhat large subleading
corrections. This is why in our original paper in Ref.
~\cite{Knecht} we decided to take a short detour. The operator
$Q_7$ mixes through the renormalization group into $Q_8$; this is
tantamount to saying that $<O_2>$ controls the large-$Q^2$
fall-off of $\Pi_{LR}(Q^2)$. Consequently, by expanding Eq.
(\ref{PiMHA}) and selecting the coefficient in $1/Q^6$, one
obtains
\begin{eqnarray}\label{matching}
  - Q^6 \Pi_{LR}(Q^2) &\longrightarrow_{_{\!\!\!\!\!\!\!\!\!\!\!\!\!\!\!\!
   Q^2\rightarrow
  \infty}}& f_0^2 M_V^2 M_A^2\qquad  (\mathrm{from\ Eq. (\ref{PiMHA})})\nonumber \\
- Q^6
\Pi_{LR}(Q^2)&\longrightarrow_{_{\!\!\!\!\!\!\!\!\!\!\!\!\!\!\!\!
Q^2\rightarrow \infty}}& 16 \pi \alpha_s
  \left(1 + \xi \frac{\alpha_s}{\pi}\right)
  < O_2(\mu) >+  \cdots \quad (\mathrm{from\ the \ OPE})\ ,
\end{eqnarray}
where $\xi=(25/8,21/8)$ in the (NDR, HV)
schemes~\cite{Donoghue,Bijnens}, and the ellipses stand for
numerically negligible terms~\cite{Knecht}\footnote{At the time
we wrote our paper ~\cite{Knecht} the term proprtional to $\xi$
was not known.}. Inputting the value of $\alpha_{s}(2\
\mathrm{GeV})\simeq 0.33 \pm 0.04$ one easily obtains $<O_2(2\
\mathrm{GeV})>$ from matching the two asymptotic behaviors in Eq.
(\ref{matching}). Saturation of this value for $<O_2(2\
\mathrm{GeV})>$ by the quark condensate in Eq. (\ref{O2}) leads to
$\langle\overline{\psi} \psi\rangle_{\overline{MS}}(2\ GeV)
\simeq - (300\pm 20\ MeV)^3$, where the error is a combined
estimate of unknown $1/N_c$ and $\alpha_{s}$ corrections.

At any rate, having the value of $<O_2(2\ \mathrm{GeV})>$ one can
easily compute any $K-\pi$ matrix element through Eq.
(\ref{Q8bos}). Table 1 shows our results together with those
obtained by other groups. A look at the central values (and
errors!) in Table 1 reveals that, even though a lot of progress
has been made, meson weak matrix elements are a very hard problem
to understand and we are not quite there, yet. Although it should
be obvious, it is worth remarking that lattice and continuum
approaches may nicely complement each other in this effort.

I would like to finish by emphasizing that the Hadronic
Approximation to large-$N_c$ QCD is not limited to simple
operators like $Q_{7,8}$ only, but is a general framework which
has also been successfully used in a variety of other problems:
$B_K$~\cite{Bk}, $\pi^0\rightarrow e^+e^-$ and $\eta \rightarrow
\mu^+\mu^-$~\cite{decay}, hadronic contributions to
$g-2$~\cite{g-2}, etc...; even in those cases when the underlying
QCD Green's function could not be related to any experimental
data --which are the most--. More work is in progress.

\begin{table}[t]
\begin{center}
\footnotesize
\begin{tabular}{|c|c|c|c|c|}
  \hline
Refs. & $M_7$(NDR) & $M_7$(HV) & $M_8$(NDR) & $M_8$(HV)
 \\\hline
$\clubsuit $(This work)  Knecht {\it et al.}\cite{Knecht}$\vphantom{X^{X^X}}$ & $0.11\pm 0.03$ & $0.67\pm0.20$ & $2.34\pm0.73$ & $2.52\pm0.79$ \\
 Narison\cite{Narison} & $0.17\pm 0.05$ &  & $1.4\pm 0.3$ &  \\
Cirigliano {\it et al.}\cite{Donoghue} & $0.16\pm0.10$ & $0.49\pm 0.07$ & $2.22\pm0.67$ & $2.46\pm 0.70$ \\
Maltman \cite{Maltman} & $0.21\pm0.03$ & $0.46\pm 0.08$ &
$1.65\pm0.45$ & $1.84\pm 0.46$\\
 Bijnens {\it et al.}\cite{Bijnens} & $0.24\pm 0.03$ & $0.37\pm 0.08$ & $1.2\pm 0.9$ & $1.3\pm 0.9$ \\
 Battacharya {\it et al.}\cite{Batta} & $0.32\pm 0.06$ &  & $1.2\pm 0.2$ &  \\
 Donini {\it et al.}\cite{Donini} & $0.11\pm 0.04$ & $0.18\pm 0.06$ & $0.51\pm 0.10$ & $0.62\pm 0.12$
  \\ \hline
\end{tabular}
\end{center}
\caption{Summary of results for $M_{7,8}\equiv< \!
(\pi\pi)_{I=2}|Q_{7,8}|K^0 \! >(2\ {\rm GeV})$, in units of ${\rm
GeV}^3$.} \vspace{-0.2cm}
\end{table}

\section*{Acknowledgments}
I wish to thank M. Knecht and E. de Rafael for a very enjoyable
and fruitful collaboration which led to this work, and to M.
Golterman, M. Perrottet and T. Hambye for interesting discussions
on this or closely-related subjects. This work has been partially
supported by CICYT AEN99-0766, CICYT-FEDER  FPA2002-00748, 2001
SGR00188 and TMR, EC-Contract No. HPRN-CT-2002-00311(EURIDICE).

\end{document}